\documentclass[aps,prl,twocolumn,superscriptaddress,floatfix]{revtex4}
\pdfoutput=1
\usepackage{graphicx}
\usepackage{dcolumn}
\usepackage{bm}
\usepackage{amsfonts}
\usepackage{amsmath}
\usepackage{color}

\begin{document}

\title{Drude weight, plasmon dispersion, and pseudospin response in doped graphene sheets}
\author{Marco Polini}
\email{m.polini@sns.it}
\affiliation{NEST-CNR-INFM and Scuola Normale Superiore, I-56126 Pisa, Italy}
\author{A.H. MacDonald}
\affiliation{Department of Physics, University of Texas at Austin, Austin, Texas 78712, USA}
\author{G. Vignale}
\affiliation{Department of Physics and Astronomy, University of Missouri, Columbia, Missouri 65211, USA}

\begin{abstract}
Plasmons in ordinary electron liquids are collective excitations whose long-wavelength limit is
rigid center-of-mass motion with a dispersion relation that is, as a 
consequence of Galileian invariance, unrenormalized by many-body effects. 
The long-wavelength plasmon frequency is related by the f-sum rule to the integral of the conductivity 
over the electron-liquid's Drude peak, implying that transport properties also tend not to have 
important electron-electron interaction renormalizations.  In this article we demonstrate 
that the plasmon frequency and Drude weight of the electron liquid in a doped graphene sheet, which is described by a massless Dirac Hamiltonian 
and not invariant under ordinary Galileian boosts, are strongly renormalized even in the long-wavelength limit. 
This effect is not captured by the Random Phase Approximation (RPA), commonly used to describe electron fluids.  It 
is due primarily to non-local inter-band exchange interactions, 
which, as we show,  {\it reduce} both the plasmon frequency and the Drude weight relative to the RPA value. 
Our predictions can be checked using inelastic light scattering or infrared spectroscopy.
\end{abstract} 

\maketitle

\section{Introduction}
The first theory of classical collective electron density oscillations in ionized gases 
by Tonks and Langmuir$^{\,}$\cite{tonks_langmuir_pr_1929} in the 1920's helped initiate the field of 
plasma physics.  The theory of collective electron density oscillations in metals, quantum in this case because of higher electron densities, 
was developed by Bohm and Pines$^{\,}$\cite{pines_bohm_pr_1952,Pines_and_Nozieres} in the 1950's and stands as a similarly pioneering contribution to many-electron physics.  Bohm and Pines coined the term {\it plasmon} to describe quantized density oscillations.    
Today {\it plasmonics} is a very active subfield of optoelectronics$^{\,}$\cite{Ebbesen_PT_2008,Maier07},  
whose aim is to exploit plasmon properties in order to compress infrared electromagnetic waves to the nanometer 
scale of modern electronic devices.  This wide importance of plasmons across different fields of 
basic and applied physics follows from the ubiquity of 
charged particles and from the strength of the long-range Coulomb interaction.  

The physical origin of plasmons is very simple. When electrons in a plasma move to screen a charge inhomogeneity, they tend to overshoot the mark. 
They are then pulled back toward the charge disturbance and overshoot again, setting up a weakly damped oscillation.
The restoring force responsible for the oscillation is the average self-consistent field created by all the electrons. Because of the long-range nature of the Coulomb interaction, the frequency of oscillations $\omega_{\rm pl}(q)$ tends to be high and is given in the long wavelength limit by $\omega^2_{\rm pl}(q\to 0)= n q^2 V_q/m$ where $n$ is the electron density, $m$ is the bare electron mass in vacuum, and $V_q$ is the Fourier transform of the Coulomb interaction.  This simple explicit plasmon energy expression 
is exact because long-wavelength plasmons involve rigid motion of the entire plasma which is 
independent of the complex exchange and correlation effects that dress$^{\,}$\cite{Giuliani_and_Vignale} the motion of an individual electron. 
The exact plasmon frequency expression is correctly captured by the celebrated RPA$^{\,}$\cite{Pines_and_Nozieres,pines_bohm_pr_1952,Giuliani_and_Vignale}, but also by rigorous arguments$^{\,}$\cite{morchio_strocchi_ap_1986} in which the selection of a particular center-of-mass position 
breaks the system's Galilean invariance and 
plasmon excitations play the role of Goldstone bosons.  In 
two-dimensional (2D) systems $V_q =2\pi e^2/q$ so that $\omega_{\rm pl}(q\to 0) = \sqrt{2\pi n e^2q/m}$.

Electrons in a solid, unlike electrons in a plasma or electrons with a {\it jellium model}$^{\,}$\cite{Giuliani_and_Vignale} background,
experience a periodic external potential created by the ions which breaks translational invariance and hence also Galilean invariance.
Solid state effects can lead in general to a renormalization of the plasmon frequency, or even to the absence of sharp plasmonic excitations. 
In semiconductors and semimetals, however, electron waves can be described at super-atomic length scales using ${\bm k} \cdot {\bm p}$ theory$^{\,}$\cite{Cardona}, which is based on an expansion of the crystal's Bloch Hamiltonian around band extrema.  In the simplest case, for example for the conduction band of common cubic semiconductors, this device leads us back to a Galileian-invariant parabolic band continuum model with isolated electron energy 
$E_{\rm c}({\bm p}) = {\bm p}^2/(2m_{\rm b})$.  The crystal background for electron waves appears only via the replacement of the 
bare electron mass by an effective band mass $m_{\rm b}$.  It is this type of ${\bm k} \cdot {\bm p}$ Galilean invariant interacting 
electron model, valid for many semiconductor and semiconductor heterojunction systems, which has been of greatest interest in solids.  
The absence of electron-electron interaction corrections to plasmon frequencies at very long wavelengths in these systems 
has been amply demonstrated experimentally by means of inelastic light scattering$^{\,}$\cite{vittorio_ils,Hirjibehedin_prb_2007}.

The situation turns out to be quite different in graphene -- 
a monolayer of carbon atoms tightly packed in a 2D honeycomb lattice$^{\,}$\cite{geim_novoselov_nat_mat_2007,allan_pt_2007,castro_neto_rmp_2009}, 
which has engendered a great deal of interest because of the new physics it exhibits 
and because of its potential as a new material for electronic technology. 
The agent responsible for many of 
the interesting electronic properties of graphene sheets is the bipartite nature of its honeycomb lattice. 
The two inequivalent sites in the unit cell of this lattice are analogous to the two spin orientations of a spin-$1/2$ particle along the $+{\hat {\bm z}}$ and $-{\hat {\bm z}}$ directions (the ${\hat {\bm z}}$ axis being perpendicular to the graphene plane).  This observation opens the way to an elegant description of electrons in graphene as particles endowed with a {\it pseudospin} 
degree-of-freedom$^{\,}$\cite{geim_novoselov_nat_mat_2007,allan_pt_2007,castro_neto_rmp_2009} 
(in addition to the regular spin degree-of-freedom which plays a passive role here).   
When ${\bm k} \cdot {\bm p}$ theory is applied to graphene it leads to a new type of electron fluid model, one with separate Dirac-Weyl Hamiltonians for 
electron waves centered in momentum space on one of two honeycomb lattice Brillouin-zone corners: 
${\hat {\cal H}}_{\rm D} = v {\bm \sigma} \cdot {\bm p}$.  Here $v$ is the bare electron velocity, ${\bm p}$ is the ${\bm k} \cdot {\bm p}$ momentum, 
and ${\bm \sigma}$ is the pseudospin operator constructed with two Pauli matrices $\{\sigma^i,i=x,y\}$, which act on the sublattice pseudospin degree-of-freedom. It follows that the energy eigenstates for a given ${\bm p}$ have pseudospins oriented either parallel (upper band) or antiparallel (lower band) to ${\bm p}$.  Physically, the orientation of the pseudospin determines the relative amplitude and the relative phase of electron waves on the two distinct graphene sublattices. 

\begin{figure}[t]
\tabcolsep=0cm
\begin{tabular}{cc}
\includegraphics[width=0.50\linewidth]{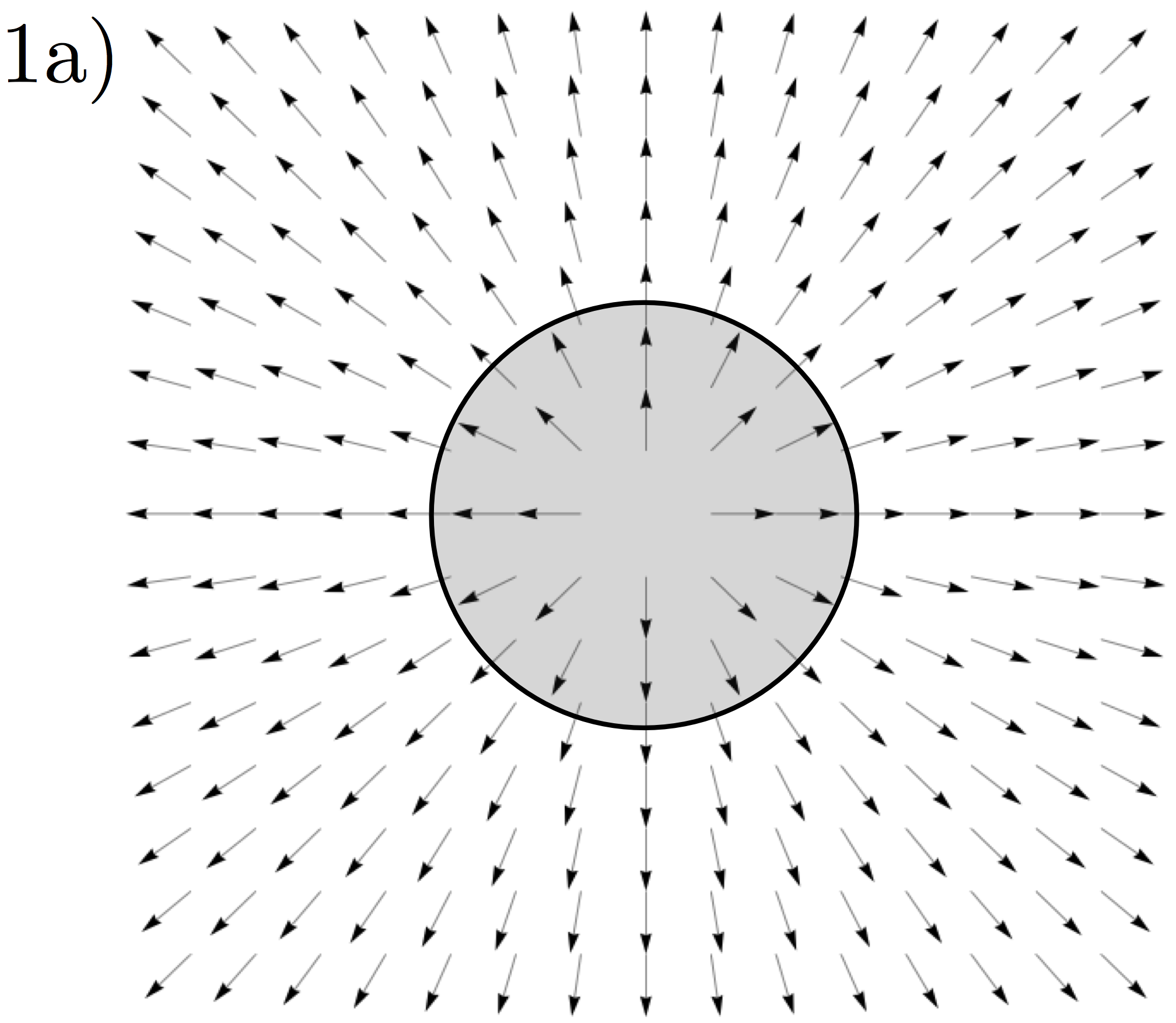} &
\includegraphics[width=0.50\linewidth]{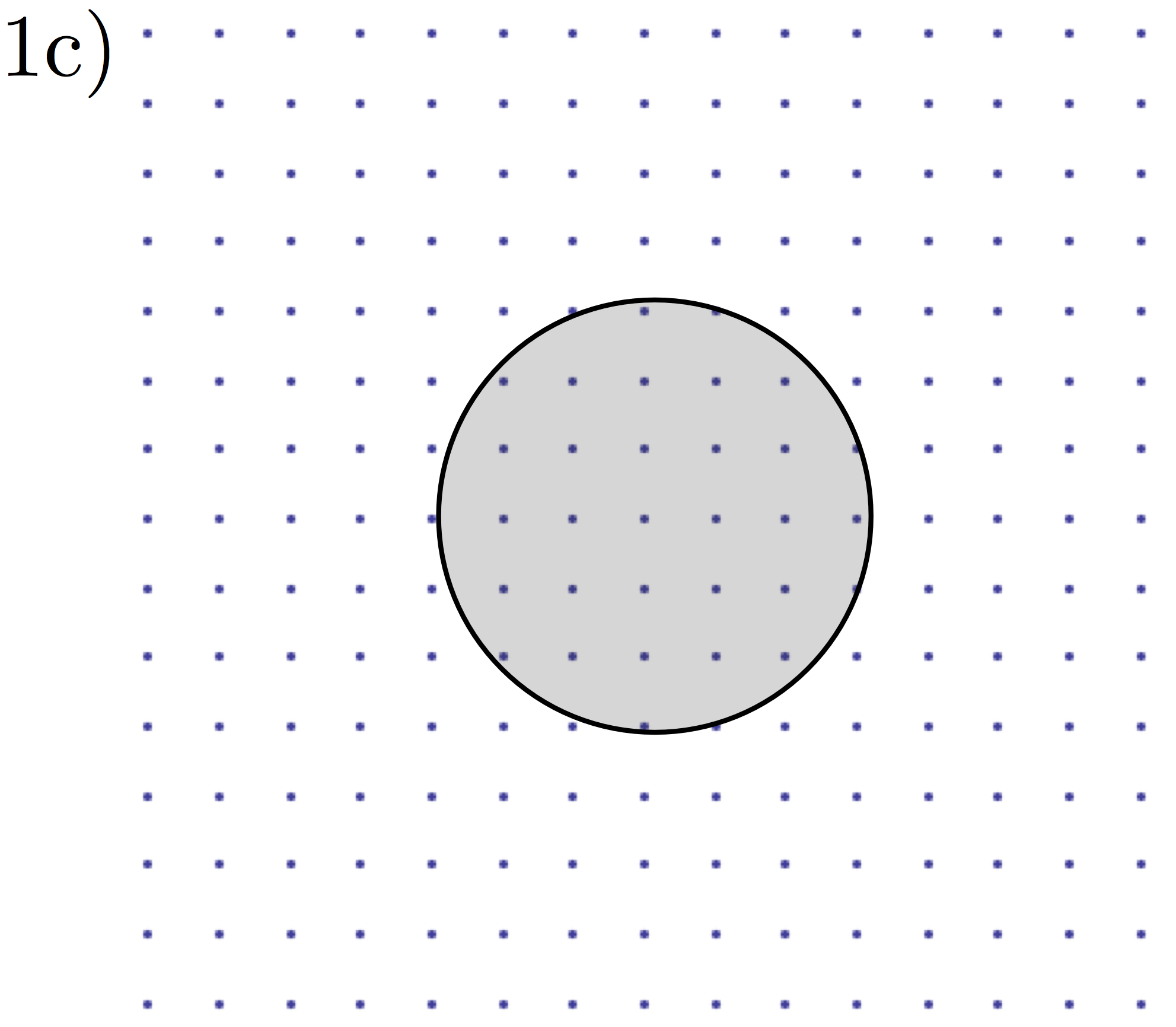}\\
\includegraphics[width=0.50\linewidth]{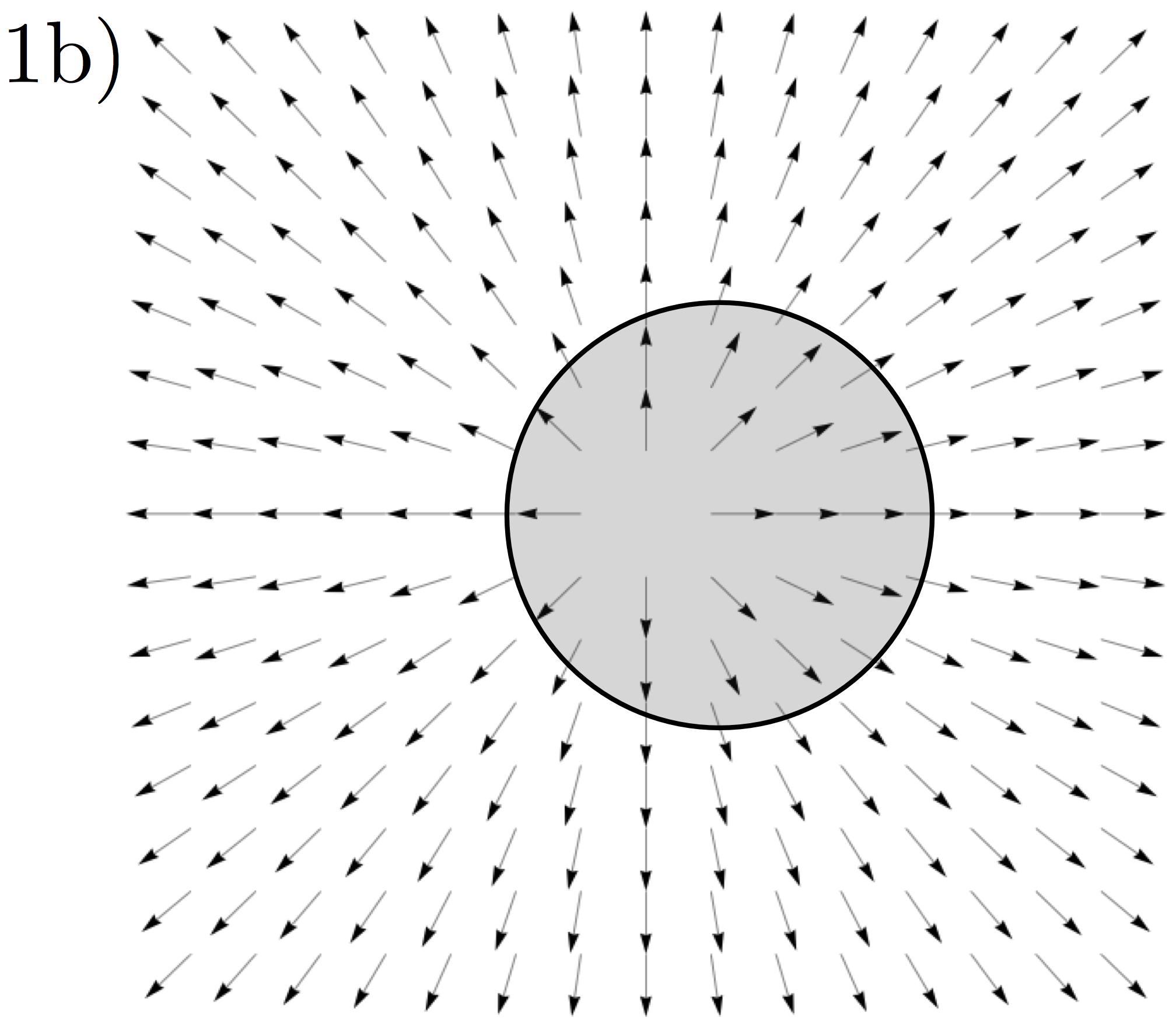} &
\includegraphics[width=0.50\linewidth]{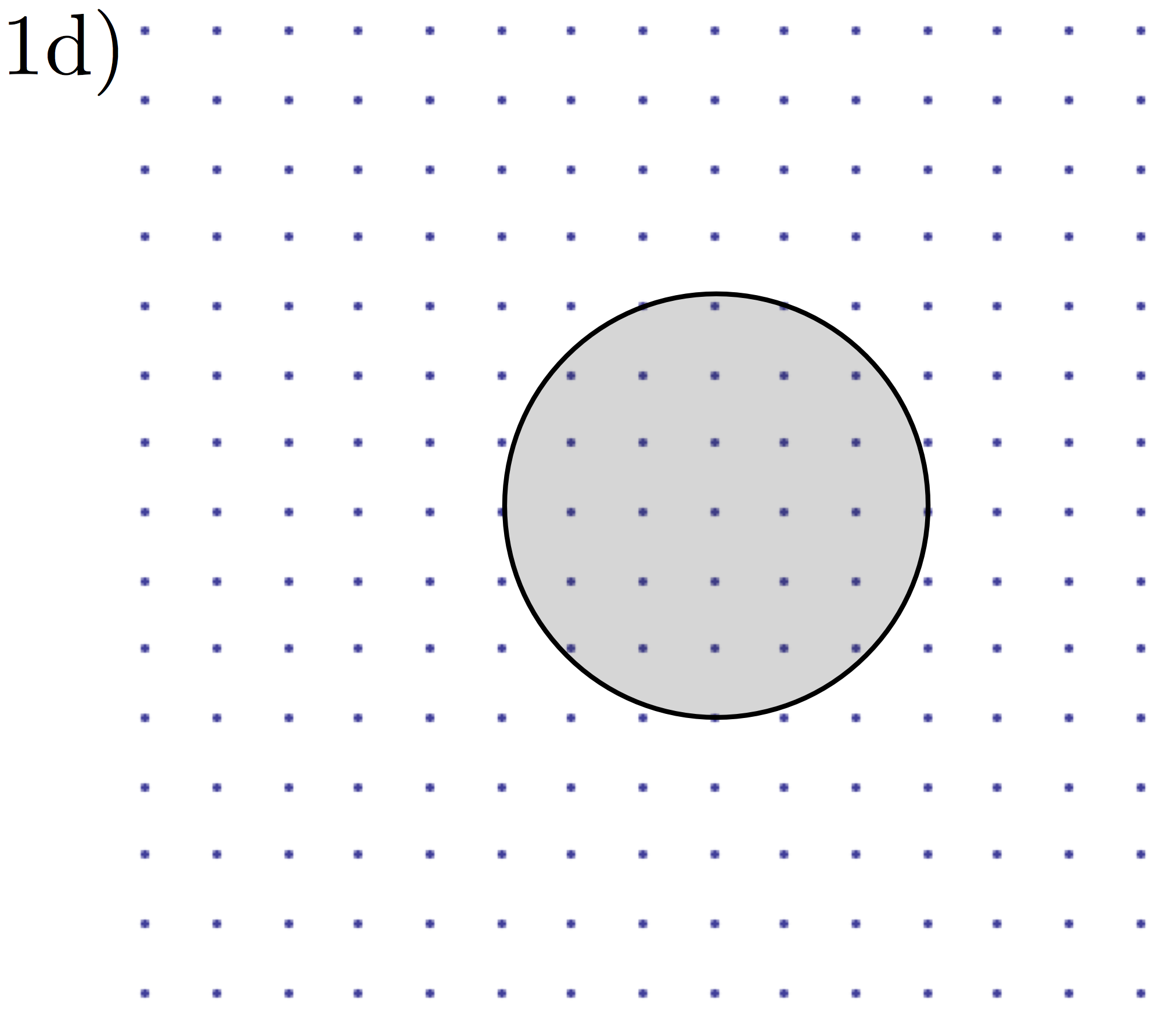}\\
\end{tabular}
\caption{{\bf Breakdown of Galileian invariance in graphene}. Panel 1a) shows the occupied electronic states in the upper band of graphene in the ground state.  Notice that every state is characterized by a value of momentum (the origin of the arrow) and a pseudospin orientation (the direction of the arrow).  Panel 1b) shows the occupied states {\it after} a Galilean boost.  An observer riding along with the boost would clearly see that the orientation of the pseudospins, relative to the center of the occupied region has changed.  It looks like the pseudospins are subjected to a ``pseudomagnetic field" that causes them to tilt towards the $+{\hat {\bm x}}$ direction.  The appearance of this pseudomagnetic field is the signature of broken Galilean invariance.  In contrast, in a Galilean invariant system [Panels 1c) and 1d)] the energy eigenstates are characterized by momentum only:  an observer riding along with the boost would not see any change in the character of the occupied states.\label{fig:one}}
\end{figure}

The feature of graphene that is ultimately responsible for the large many-body effects on the plasmon dispersion and the Drude weight is 
{\it broken Galilean invariance}.  What happens is that the oriented  pseudospins provide an ``ether" against which a global boost of the momenta becomes detectable. This is explained in detail in the caption of Fig.~\ref{fig:one}.

From Fig.~\ref{fig:one} we can also see why the plasmon frequency in graphene is so strongly affected by exchange and 
correlation.  In a plasmon mode the region of occupied states (Fermi circle) oscillates back and forth in momentum space under the action of the self-induced electrostatic field.  In graphene however, this oscillatory motion is inevitably coupled with an oscillatory motion of the pseudospins.  Since exchange interactions depend on the relative orientation of pseudospins they contribute to plasmon kinetic energy and renormalize the plasmon frequency even at leading order in $q$.

In what follows we present a many-body theory of this subtle pseudospin coupling effect and 
discuss the main implications of our findings for theories of charge transport and collective 
excitations in doped graphene sheets.

\section{Graphene Dirac Model}
\label{sect:model}

Graphene's honeycomb lattice has two-atoms per unit 
cell and its $\pi$-valence band and $\pi^*$-conduction band touch at two inequivalent 
points, $K$ and $K'$, in the honeycomb lattice Brillouin-zone.  The energy bands near {\it e.g.} 
the $K$ point are described at low energies by the spin-independent massless Dirac Hamiltonian ${\hat {\cal H}}_{\rm D}$ introduced above. Electron-electron interactions in graphene 
are described by the usual non-relativistic Coulomb Hamiltonian ${\hat {\cal H}}_{\rm C}$, 
which is controlled by 
the 2D Fourier transform of the Coulomb interaction, 
$V_q=2\pi e^2/(\epsilon q)$ with $\epsilon$ an effective average dielectric constant.

Electron carriers with density $n$ can be induced in graphene by purely electrostatic means, creating a circular 2D Fermi surface 
in the conduction band with a Fermi radius $k_{\rm F}$, which is proportional to $\sqrt{n}$$^{\,}$\cite{electron_doping}. 
The model described by ${\hat {\cal H}} = {\hat {\cal H}}_{\rm D} + {\hat {\cal H}}_{\rm C}$ requires an ultraviolet wavevector cut-off,
 $k_{\rm max}$, which should be assigned a value corresponding to the
wavevector range over which ${\hat {\cal H}}_{\rm D}$ describes graphene's $\pi$ bands. This corresponds to taking 
$k_{\rm max} \sim 1/a_0$ where $a_0 \sim 1.42$~\AA~is the carbon-carbon distance. 
This model is useful when $k_{\rm max}$ is much larger than $k_{\rm F}$. 
In this low-energy description, the many-body properties of doped graphene sheets depend$^{\,}$\cite{barlas_prl_2007,polini_ssc_2007} 
on the dimensionless fine-structure coupling constant 
$\alpha_{\rm ee}= e^2/(\epsilon \hbar v)$ (which is defined as the ratio between the Coulomb energy scale $e^2 k_{\rm F}/\epsilon$ and the kinetic energy scale $\hbar v k_{\rm F}$) and on density {\it via} the ultraviolet cut-off $\Lambda = k_{\rm max}/k_{\rm F}$. 
The fine-structure constant $\alpha_{\rm ee}$ can be tuned experimentally 
by changing the dielectric environment surrounding the graphene flake$^{\,}$\cite{jang_prl_2008,mohiuddin_preprint_2008}. 
The ultraviolet cut-off $\Lambda$ varies from $\sim 10$ for a very high-density graphene system with $n \sim 10^{14}~{\rm cm}^{-2}$
to $\sim 100$ for a density $n \sim 10^{12}~{\rm cm}^{-2}$ just large enough to screen out unintended$^{\,}$\cite{yacoby_natphys_2008} inhomogeneities.
From now on Planck's constant $h$ divided by $2\pi$ will be set equal to unity, $\hbar=1$.

The collective (plasmon) modes of the system can be found by solving the following equation$^{\,}$\cite{Giuliani_and_Vignale},
\begin{equation}\label{eq:plasmon_equation}
1 - V_q {\widetilde \chi}_{\rho\rho}(q,\omega) = 0~
\end{equation}
where ${\widetilde \chi}_{\rho\rho}(q,\omega)$ is the so-called {\it proper}$^{\,}$\cite{proper_diagrams} 
density-density response function. In the $q \to 0$ limit of interest here we can neglect the distinction between the proper and 
the full causal response function $\chi_{\rho\rho}(q,\omega)$.  We show below that
\begin{equation}\label{eq:calA}
\lim_{\omega \to 0} \lim_{q \to 0} \Re e~\chi_{\rho\rho}(q,\omega) = 
{\cal A}~\frac{v^2 q^2}{\omega^2}~
\end{equation}
where ${\cal A}$ is a density-dependent constant which has the value  
${\cal A}_0= g \varepsilon_{\rm F}/(4\pi v^2)$ for $\alpha_{\rm ee} \to 0$. 
Here $g = g_{\rm s} g_{\rm v}=4$ accounts for spin and valley degeneracy and $\varepsilon_{\rm F}=v k_{\rm F}$ is the Fermi energy.  
Note the order of limits in Eq.~(\ref{eq:calA}) and below: the limit $\omega \to 0$ is always taken in the dynamical sense, {\it i.e.} 
$v q \ll \omega \ll 2\varepsilon_{\rm F}$. Using Eq.~(\ref{eq:calA}) in Eq.~(\ref{eq:plasmon_equation}) and solving for $\omega$ we find that, to leading order in $q$, 
\begin{equation}\label{MainResult}
\omega^2_{\rm pl}(q\to 0)= \frac{2 \pi e^2 v^2 {\cal A}}{\epsilon}~q~.
\end{equation}
In the same limit the imaginary-part of the low-frequency conductivity $\sigma(\omega) = i e^2 \omega \chi_{\rho\rho}(\omega)/q^2$ has the form
\begin{equation} \label{DrudeWeightForm} 
\Im m~\sigma(\omega) \to \frac{e^2 v^2 {\cal A}}{\omega}~.
\end{equation} 
It then follows from a standard Kramers-Kr\"onig analysis that the real-part of the conductivity has a $\delta$-function peak at $\omega=0$:
$\Re e~\sigma(\omega) = {\cal D} \delta(\omega)$ where the Drude weight
\begin{equation} \label{DrudeWeight} 
{\cal D} = \pi e^2 v^2 {\cal A}~.
\end{equation}
In the presence of disorder the $\delta$-function peak is broadened into a Drude peak, but the Drude weight is preserved. 
The Drude weight ${\cal D}$ defines an effective f-sum rule (cf. Ref.$^{\,}$\cite{sabio_prb_2008}) 
in the dynamical regime $v q \ll \omega \ll 2\varepsilon_{\rm F}$$^{\,}$\cite{low_frequency}. 

We thus see from Eqs.~(\ref{MainResult}) and~(\ref{DrudeWeight}) 
that the quantity ${\cal A}$ completely controls the plasmon dispersion at long wavelengths and the Drude weight. 
In the following section we first relate ${\cal A}$ to the longitudinal pseudospin susceptibility and then carry 
out a self-consistent microscopic calculation of ${\cal A}$ 
which demonstrates that its value is suppressed by electron-electron interactions. When this renormalization is neglected ${\cal A} \to {\cal A}_0$ and  
\begin{equation}\label{eq:RPA}
\omega^2_{\rm pl}(q \to 0)= \varepsilon^2_{\rm F}~\frac{g\alpha_{\rm ee}}{2}~\frac{q}{k_{\rm F}}~,
\end{equation}
the RPA$^{\,}$\cite{wunsch_njp_2006,hwang_prb_2007,polini_prb_2008} result for the plasmon dispersion at long wavelengths. 
The RPA Drude weight ${\cal D}= 4 \varepsilon_{\rm F} \sigma_{\rm uni}$ where, restoring Planck's constant for a moment, 
$\sigma_{\rm uni}= e^2/(4\hbar)$ is the so-called 
universal$^{\,}$\cite{nair_science_2008,wang_science_2008,li_natphys_2008,mak_prl_2008} frequency-independent interband conductivity
of a neutral graphene sheet. 
We show that both $\omega_{\rm pl}(q\to 0)$ and ${\cal D}$ are substantially altered by electron-electron interactions. 

\section{Pseudospin Response and Dirac-Model Plasmons}
\label{sect:eof}

By using the equation of motion for the density operators$^{\,}$\cite{Giuliani_and_Vignale} which appear in the  
density-density response function, $\chi_{\rho\rho}(q,\omega)$ can be reexpressed in terms of the longitudinal current-current response function.  
When this procedure is applied to a Galilean-invariant system with mass $m$ it leads immediately to the well-known result 
${\displaystyle \lim_{\omega \to 0} \lim_{q \to 0} \Re e~\chi_{\rho\rho}(q,\omega) = nq^2/(m\omega^2)}$.  
In the case of graphene, however, the current operator (defined as the derivative of the Hamiltonian with respect to ${\bm k}$) 
is directly proportional to the pseudospin operator$^{\,}$\cite{katsnelson_ssc_2007} and we obtain instead
\begin{equation}\label{eq:eom_mdf}
\chi_{\rho\rho}(q,\omega) =
\frac{vq}{\omega^2}  \langle [{\hat \sigma}^x_{\bm q}, {\hat \rho}_{-{\bm q}}]\rangle
+\frac{v^2q^2}{\omega^2}\chi_{\sigma^x \sigma^x}(q,\omega)~,
\end{equation}
where ${\hat \sigma}^x_{\bm q}$ is the component of the pseudospin fluctuation operator along the direction of ${\bm q}$, which we assume to be the ${\hat {\bm x}}$ direction, and $\chi_{\sigma^x \sigma^x}(q,\omega)$ is the longitudinal pseudospin-pseudospin response function. The latter describes the response of ${\hat \sigma}^x_{\bm q}$ to a pseudomagnetic field $B_{\bm q}$ which enters the Hamiltonian with a term of the form 
${\hat \sigma}^x_{-{\bm q}}B_{\bm q}$ (notice that this has the opposite sign compared to the usual Zeeman coupling).
 
Because of the presence of the infinite sea of negative energy states Eq.~(\ref{eq:eom_mdf}) must be handled with great care$^{\,}$\cite{anomalous_commutator}. In the noninteracting case we have
\begin{equation}\label{eq:noninteracting_anomalous_commutator}
\langle[{\hat \sigma}^x_{\bm q}, {\hat \rho}_{-{\bm q}}]\rangle = \sum_{|{\bm k}| < \Lambda k_{\rm F}}
[\cos(\varphi_{\bm k})n^{(0)}_{{\bm k}, -}-\cos(\varphi_{{\bm k}-{\bm q}})n^{(0)}_{{\bm k}-{\bm q},-}]~, 
\end{equation}
where $\varphi_{\bm k}$ is the angle between ${\bm k}$ and the ${\hat {\bm x}}$ axis and $n^{(0)}_{{\bm k}, -}=1$ is the occupation of the lower band. 
For $q \ll \Lambda k_{\rm F}$ this can be rewritten as the sum of $\cos(\varphi_{\bm k})$ over the region comprised between the circles $|{\bm k}| < \Lambda k_{\rm F}$ and $|{\bm k}-{\bm q}| < \Lambda k_{\rm F}$: 
only states deep in the negative energy Dirac sea contribute. When interactions are included $n^{(0)}_{{\bm k}, -}$ is replaced by exact occupation numbers and additional terms associated with pseudospin-orientation fluctuations appear. In the next section we present a microscopic time-dependent Hartree-Fock theory of ${\cal A}$ that is valid to first order in $\alpha_{\rm ee}$. Since this theory neglects ground-state pseudospin and occupation-number fluctuations, which are of second order in $\alpha_{\rm ee}$,  the anomalous commutator can be consistently evaluated in the noninteracting electron ground state and we find that$^{\,}$\cite{sabio_prb_2008}
\begin{equation}\label{eq:anomalous_commutator_final}
\lim_{q \to 0}~\frac{v q}{\omega^2}~\langle[{\hat \sigma}^x_{\bm q}, {\hat \rho}_{-{\bm q}}]\rangle = 
\frac{v^2q^2}{\omega^2}\frac{g \varepsilon_{\rm max}}{4\pi v^2}~.
\end{equation}
As we explain below, $g \varepsilon_{\rm max}/(4\pi v^2)$ is the negative of the pseudospin susceptibility of a {\it noninteracting undoped} graphene sheet, 
{\it i.e.}
\begin{equation}\label{eq:claim_proven_below}
\frac{g \varepsilon_{\rm max}}{4\pi v^2} = - \lim_{\omega \to 0}\lim_{q\to 0} \Re e~\chi^{(0{\rm u})}_{\sigma^x\sigma^x}(q,\omega)~,
\end{equation}
where $\chi^{(0{\rm u})}_{\sigma^x\sigma^x}(q,\omega)$ is the pseudospin-pseudospin response function of the noninteracting undoped system.
Combining the two terms on the r.h.s. of Eq.~(\ref{eq:eom_mdf}) we arrive at the following expression for ${\cal A}$:
\begin{equation}\label{eq:calA_def}
{\cal A} \equiv \lim_{\omega \to 0}\lim_{q\to 0} \Re e~\left[
\chi_{\sigma^x\sigma^x}(q,\omega) -
\chi^{(0{\rm u})}_{\sigma^x\sigma^x}(q,\omega) 
\right]~.
\end{equation}
Thus ${\cal A}$ has a very clear physical meaning: it is the pseudospin susceptibility of the interacting system regularized by
subtracting the pseudospin susceptibility of the {\it reference} noninteracting undoped system.

On quite general grounds it is possible to express the fully interacting value of ${\cal A}$ in terms of a small set of dimensionless parameters by adapting to doped graphene sheets the original macroscopic phenomenological theory of Landau$^{\,}$\cite{Giuliani_and_Vignale}, which is usually applied to normal Fermi liquids. In what follows, however, we will present a microscopic theory of ${\cal A}$, which we believe to be accurate at weak coupling and which enables us to draw quantitative conclusions on the impact of electron-electron interactions on the plasmon dispersion and 
Drude weight of doped graphene sheets.

\section{Self-consistent Hartree-Fock mean-field theory of the pseudospin susceptibility}

We now proceed to a quantitative microscopic calculation of the pseudospin susceptibility ${\cal A}$ which goes beyond the RPA.   Specifically, we will take into account exactly the self-consistent exchange field which accompanies pseudospin polarization. To this end we set up the time-dependent Hartree-Fock (HF) theory of the response of the system to a uniform pseudomagnetic field  ${\bm B}_{\rm ext} = B_{\rm ext} {\hat {\bm x}}$ oriented along the ${\hat {\bm  x}}$ direction.
   
It is important to realize that the zero-frequency limit of the uniform pseudospin susceptibility is singular in the following sense.  
When $B_{\rm ext}$ is truly time-independent then its effect is simply to shift the occupied states  in ${\bm k}$-space while reorienting the pseudospins. 
Changes due to ${\bm k}$-state repopulation and pseudospin reorientation at a given ${\bm k}$ cancel each other as required by gauge invariance, since a constant $B_{\rm ext}$ can be eliminated by a gauge transformation$^{\,}$\cite{cut_off}. 
At finite frequency, however, and no matter how small the frequency, the states are unable to repopulate (${\bm k}$ is a constant of the motion in the presence of the perturbing field)  and pseudospin reorientation is the only effect that is left.  This leads to a finite regularized pseudospin response in the zero-frequency limit.  And this is clearly the limit of interest here$^{\,}$\cite{isothermal_adiabatic}.

The HF theory is usually described as an approximate factorization of the two-body interaction Hamiltonian 
into a product of  simpler one-body terms.  
In the present case, the total HF Hamiltonian can be written in the following physically transparent form:
\begin{equation}
{\hat {\cal H}}_{\rm HF} = \sum_{{\bm k}, \alpha, \beta} {\hat \psi}^\dagger_{{\bm k}, \alpha} 
[\delta_{\alpha\beta}B_0({\bm k}) +{\bm \sigma}_{\alpha\beta} \cdot {\bm B}({\bm k}) ]{\hat \psi}_{{\bm k}, \beta}~,
\end{equation}
where the HF fields $B_0({\bm k})$ and ${\bm B}({\bm k})$ are defined by
\begin{equation}\label{eq:bnot}
B_0({\bm k}) = - \int \frac{d^2{\bm k}'}{(2\pi)^2} V_{{\bm k}-{\bm k}'} f_+(k')~,
\end{equation}
and
\begin{eqnarray}\label{eq:hf-b-field}
{\bm B}({\bm k}) = B_{\rm ext}{\hat {\bm x}}+ v{\bm k}- 
\int \frac{d^2{\bm k}'}{(2\pi)^2} V_{{\bm k}-{\bm k}'} f_-(k') {\hat {\bm n}}({\bm k}')\,,
\end{eqnarray}
with $f_\pm(k)= (n^{(0)}_{{\bm k}, +} \pm n^{(0)}_{{\bm k}, -})/2$, where the $n^{(0)}_{{\bm k}, \lambda}$ are noninteracting band occupation factors,  and ${\hat {\bm n}}({\bm k}) = {\bm B}({\bm k})/|{\bm B}({\bm k})|$ is the unit vector in the direction of ${\bm B}({\bm k})$.  We have taken the limit $q \to 0$ in Eq.~(\ref{eq:hf-b-field}) by considering a spatially homogeneous external 
pseudomagnetic field applied along the ${\hat {\bm x}}$ direction.  
According to the previous discussion the occupation factors in  ${\bm k}$-space are not affected by the perturbation. 
The second term in this equation 
is the band pseudomagnetic field (see ${\hat {\cal H}}_{\rm D}$), and the last term is the exchange field. The Hamiltonian ${\hat {\cal H}}_{\rm HF}$ has two bands with energies $\varepsilon^{(\pm)}_{\rm HF}({\bm k})=B_0({\bm k}) \pm |{\bm B}({\bm k})|$.  In time-dependent HF theory electrons respond to the external field and to the induced change in the exchange field. 

In the absence of the external field ${\hat {\bm n}}({\bm k}) = {\hat {\bm k}}$, so that there is no total pseudospin polarization, 
and $|{\bm B}({\bm k})| \to |{\bm B}^{\rm eq}({\bm k})|$ depends only on $k=|{\bm k}|$~$^{\,}$\cite{polini_ssc_2007}. Our aim here is to calculate the additional exchange field that arises from the polarization of the pseudospin when $B_{\rm ext} \ne 0$, since
this determines the exchange correction to the pseudospin susceptibility.  
\begin{figure}[t]
\includegraphics[width=1.0\linewidth]{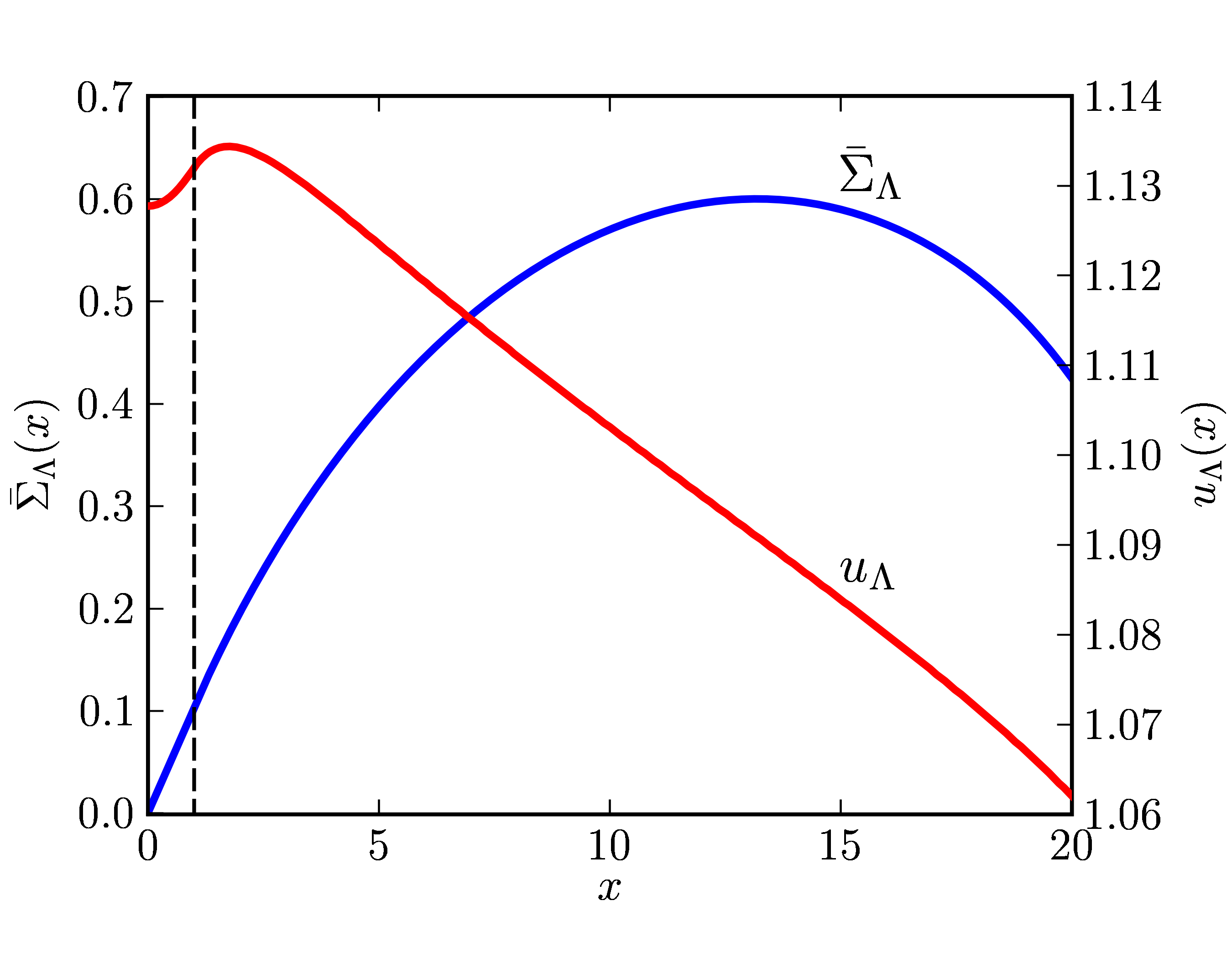}
\caption{{\bf The exchange contribution to the equilibrium pseudomagnetic field 
and the induced transverse pseudospin field}. The data shown in this figure (as well as those reported in Fig.~\ref{fig:three}) have been obtained using a Thomas-Fermi screened potential $V_q = 2\pi e^2/[\epsilon(g \alpha_{\rm ee} k_{\rm F} + q)]$. The data labeled by the blue solid line 
refer to the quantity ${\bar \Sigma}_\Lambda(x)$ plotted as a function of $x=k/k_{\rm F}$ (for wavevectors $k$ up to the ultraviolet cut-off $k_{\rm max}$) for 
$\alpha_{\rm ee}=0.2$ and $n \sim 2 \times 10^{13}~{\rm cm}^{-2}$ ($\Lambda=20$). The data labeled by the red solid line refer 
the solution $u_\Lambda(x)=\delta B_{{\rm T},1}/B_{\rm ext}$ of Eq.~(\ref{eq:integral_equation}) plotted 
as a function of $x=k/k_{\rm F}$ for the same physical 
parameters. The dashed vertical line indicates the point $k=k_{\rm F}$. Note that $u_\Lambda$ is larger than unity, thus giving an induced transverse pseudospin field $\delta B_{{\rm T},1}$ that is {\it larger} than the bare external field $B_{\rm ext}$.\label{fig:two}}
\end{figure}

For $q =0$ the response is due to vertical interband transitions at wavevectors with $|{\bm k}|> k_{\rm F}$; transitions  with $|{\bm k}| < k_{\rm F}$ are Pauli blocked.  
Because only the transverse  component $\delta {\bm B}_{\rm T}$ ($\propto  {\hat {\bm z}}\times {\hat {\bm k}}$) 
of the pseudospin field contributes to these matrix elements,  it is evident that only $\delta {\bm B}_{\rm T}$ 
is relevant to the calculation of the susceptibility. We find that    
\begin{equation} 
\delta {\bm B}_{\rm T}({\bm k}) = - [\delta B_{{\rm T},1}(k)\sin(\varphi_{\bm k})]~{\hat {\bm z}}\times 
{\hat {\bm k}}~,
\end{equation}
where $\delta B_{{\rm T},1}(k)/B_{\rm ext} \equiv u_\Lambda$ solves the integral equation:
\begin{eqnarray}\label{eq:integral_equation}
u_\Lambda(x) &=& 1 + \int_1^{\Lambda}dx' K(x,x')~u_\Lambda(x')
\end{eqnarray}
with
\begin{equation}
K(x,x') = \frac{1}{4}\alpha_{\rm ee}\frac{{\bar V}_0(x,x')+{\bar V}_2(x,x')}{1+{\bar \Sigma}_\Lambda(x')/x'}~.
\end{equation}
Here we have introduced the dimensionless Coulomb pseudopotentials
\begin{equation}\label{pseudopotentials}
{\bar V}_m(k,k') = \frac{\epsilon k_{\rm F}}{2 \pi e^2}\int_0^{2\pi} \frac{d\theta}{2\pi} e^{-i m \theta}~\left.V_q\right|_{q=|{\bm k}-{\bm k}'|}~,
\end{equation}
$\theta$ being the angle between ${\bm k}$ and ${\bm k}'$, and all wavevectors have been scaled with $k_{\rm F}$.  
The quantity
\begin{equation}
{\bar \Sigma}_\Lambda(x) = \frac{1}{2}\alpha_{\rm ee}\int_{1}^{\Lambda}dx' x' {\bar V}_1(x,x')
\end{equation}
is the exchange contribution to the equilibrium pseudomagnetic field ${\bm B}^{\rm eq}({\bm k})$.
Illustrative numerical results for ${\bar \Sigma}_\Lambda$ and $u_\Lambda$ [the latter as obtained from the self-consistent solution of Eq.~(\ref{eq:integral_equation})] 
for $\alpha_{\rm ee}=0.2$ and $\Lambda=20$ are shown in Fig.~\ref{fig:two}.

\begin{figure}[t]
\includegraphics[width=1.00\linewidth]{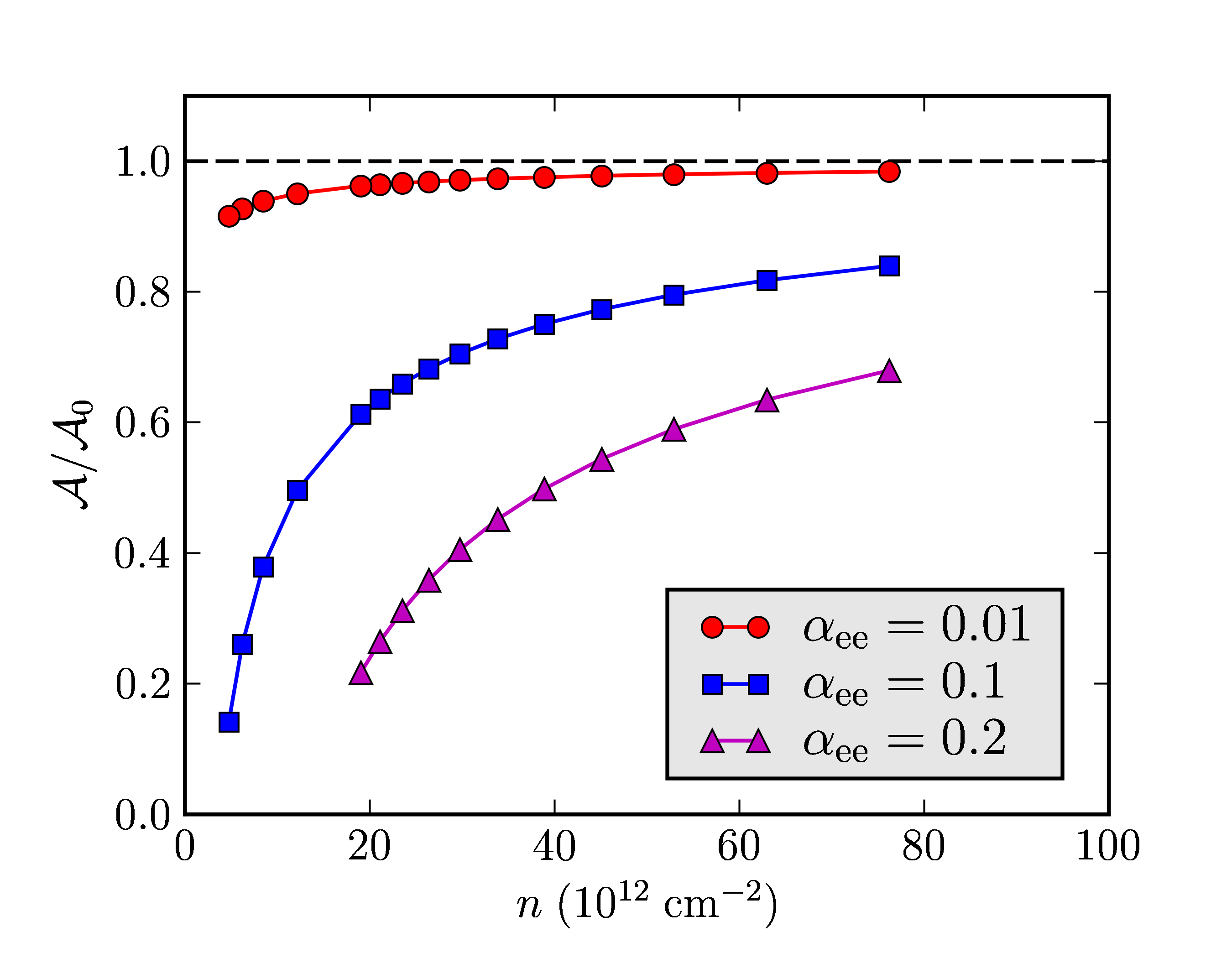}
\caption{{\bf The uniform pseudospin susceptibility of a doped graphene sheet}. 
The data labeled by filled symbols refer to the Hartree-Fock value of the ratio ${\cal A}/{\cal A}_0$, as calculated from Eq.~(\ref{eq:AHF_explicit}), 
as a function of electron density  
$n$ (in units of $10^{12}~{\rm cm}^{-2}$) for various values of graphene's fine-structure constant $\alpha_{\rm ee}$. 
The dashed horizontal line represents the prediction of the RPA$^{\,}$\cite{wunsch_njp_2006,hwang_prb_2007,polini_prb_2008}, for 
which ${\cal A}/{\cal A}_0=1$ for every value of $n$ and $\alpha_{\rm ee}$.\label{fig:three}}
\end{figure}

After straightforward manipulations we arrive at the following HF expression for the ratio between ${\cal A}$ and its noninteracting 
value ${\cal A}_0$ in terms of ${\bar \Sigma}_\Lambda(x)$ and $u_\Lambda$:
\begin{eqnarray}\label{eq:AHF_explicit}
\frac{{\cal A}}{{\cal A}_0} &=&1- 
\int_{1}^{\Lambda}dx \left[\frac{u_\Lambda(x)}{1+{\bar \Sigma}_\Lambda(x)/x}-1\right]~.
\end{eqnarray}
A plot of the ratio ${\cal A}/{\cal A}_0$ as a function of electron density for various values of $\alpha_{\rm ee}$ has 
been reported in Fig.~\ref{fig:three}.
From this plot we clearly see  that ${\cal A}/{\cal A}_0$ is substantially lower than unity. According to Eqs.~(\ref{MainResult}) and~(\ref{DrudeWeight}) this implies a strong reduction of the plasmon frequency and the Drude weight.  

In order to explain this result we show in Fig. 4 the response of the pseudospins to a pseudomagnetic field in the  $+{\hat {\bm x}}$ direction.  Notice that the pseudospins in the lower band are tilted {\it away} from the pseudomagnetic field, while those in the upper band are tilted {\it towards} the pseudomagnetic field.   Because there are many more particles in the lower band than in the upper band we see that the total pseudospin response is {\it negative}.  It is only after subtracting the noninteracting undoped response that we obtain the positive quantity  ${\cal A}$.  It should be evident from this description that any many-body effect that enhances the total pseudospin response will reduce the value of ${\cal A}$,  while  any many-body effect that suppresses the total pseudospin response will increase the value of ${\cal A}$.
\begin{figure}[t]
\tabcolsep=0cm
\begin{tabular}{c}
\includegraphics[width=0.60\linewidth]{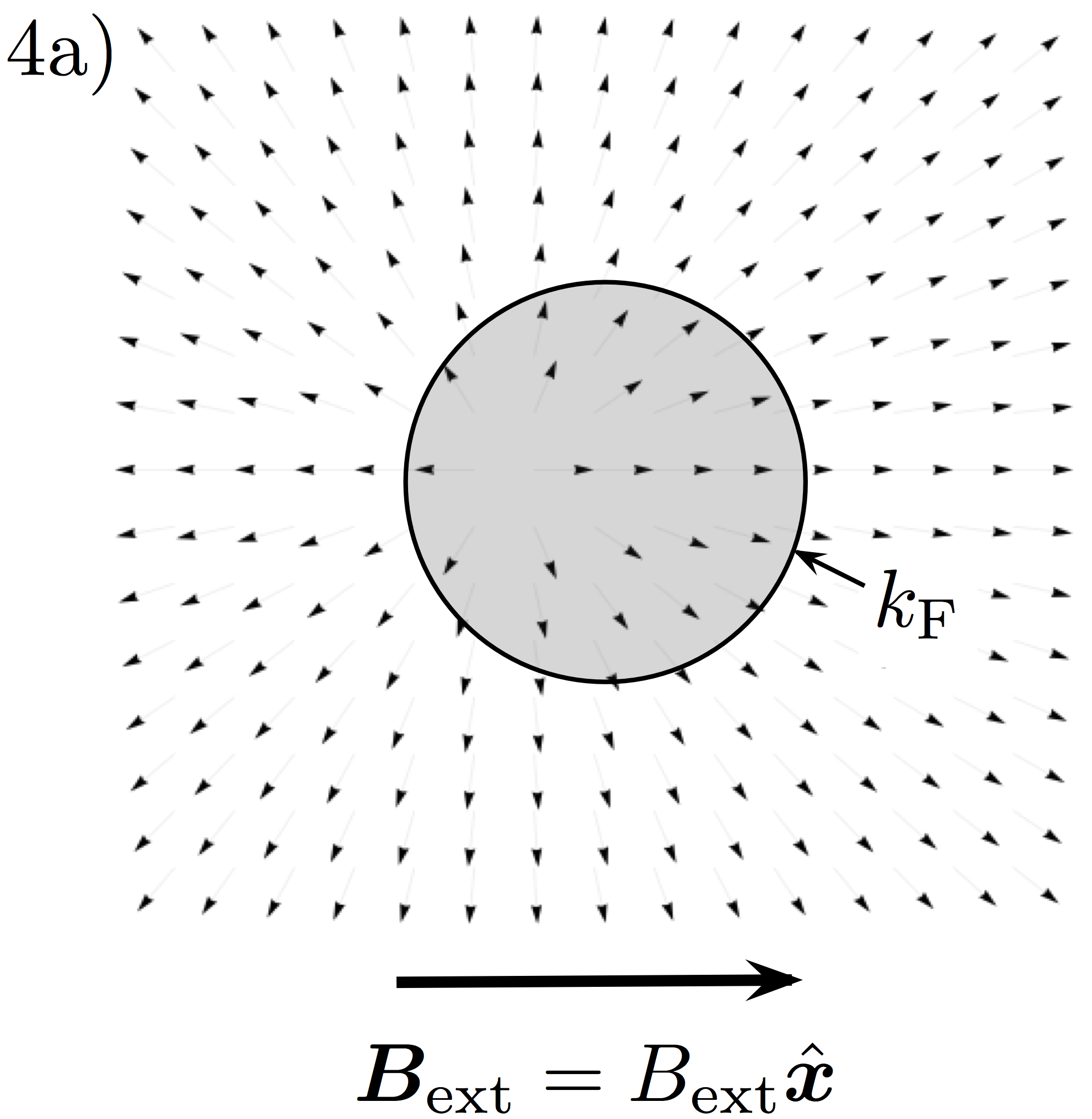} \\
\includegraphics[width=0.60\linewidth]{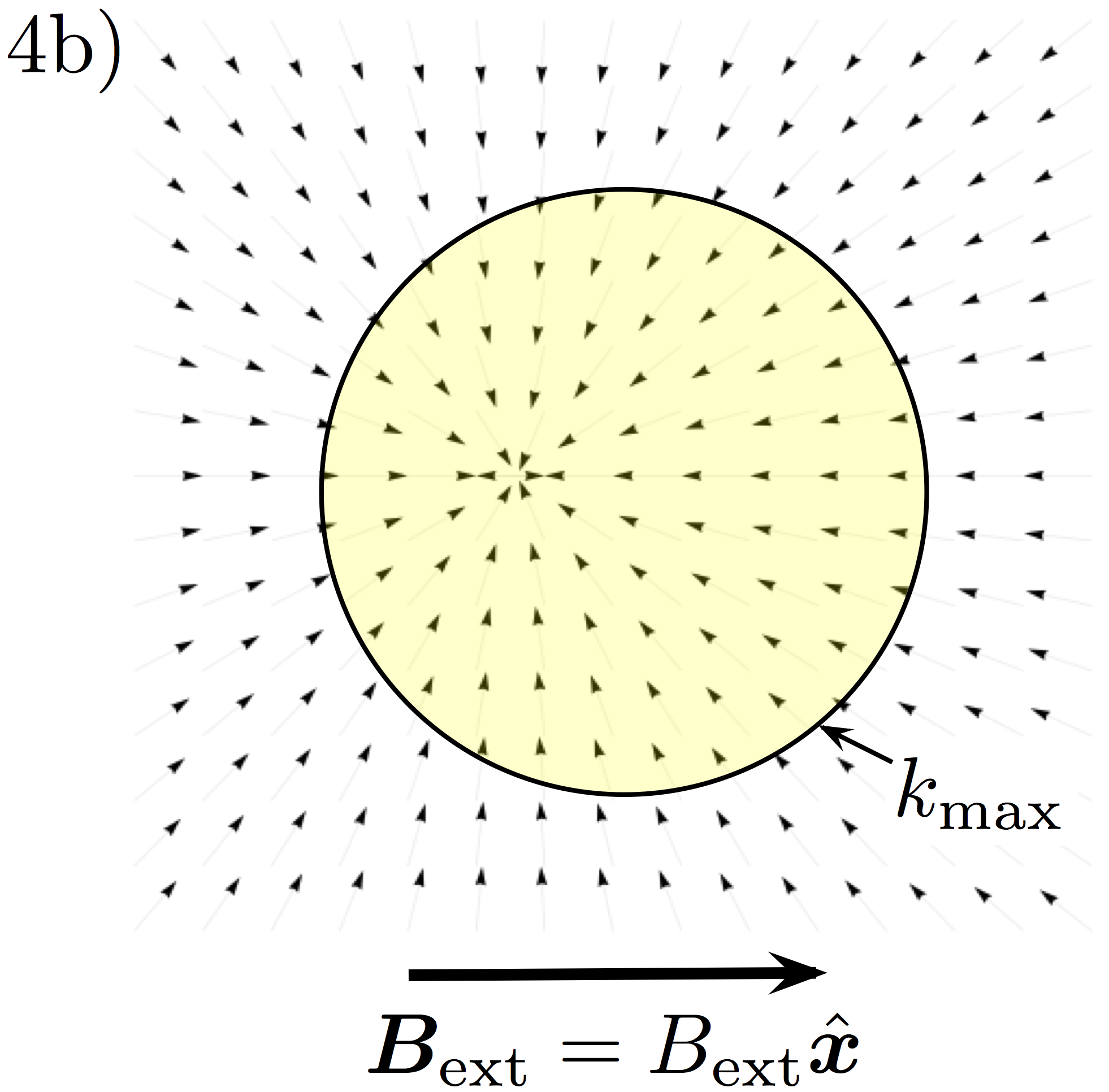}
\end{tabular}
\caption{{\bf Response of the pseudospins to a pseudomagnetic field}. Panel 4a): Pseudospins in the high-energy band tilt towards the 
pseudomagnetic field ${\bm B}_{\rm ext}$ applied along the $+{\hat {\bm x}}$ direction. Panel 4b): Pseudospins in the low-energy band tilt away from the 
pseudomagnetic field. The reason for this unusual response is easy to understand. Pseudospins in the lower band are in their ground state: because of the anomalous sign of the pseudospin-pseudomagnetic coupling mentioned in the main body of this article, they are anti-aligned with the intrinsic Dirac band pseudomagnetic field. When an additional external pseudomagnetic field is applied they will simply tilt away from it to minimize the energy. The occupied states in the higher band however, due to the Pauli principle, have pseudospins which are not in their ground state and which are aligned with the Dirac band pseudomagnetic field. When an additional external pseudomagnetic field is applied they thus respond in an unusual way tilting towards ${\bm B}_{\rm ext}$. \label{fig:four}}
\end{figure}

In the present theory interactions affect the value of ${\cal A}$ in two competing ways.  (i) The exchange field enhances $\delta B_{{\rm T}, 1}$ relative to $B_{\rm ext}$ ($u_\Lambda = \delta B_{{\rm T}, 1}/B_{\rm ext}>1$), thus enhancing the total pseudospin response.  From what we have said above it follows that this  effect gives a negative contribution to ${\cal A}$.  (ii) The exchange contribution ${\bar \Sigma}_\Lambda$ to the equilibrium pseudomagnetic field increases$^{\,}$\cite{polini_ssc_2007} the conduction valence band splitting, thus suppressing the total pseudospin response. From the previous discussion it follows that this effect gives a positive contribution to ${\cal A}$. Our calculations indicate that effect (i) dominates, resulting in a net reduction of the value of ${\cal A}$. Physically, the enhancement of the total pseudospin response (and thus the reduction of ${\cal A}$) is a consequence of the gain in exchange energy that occurs when pseudospins in the same band tilt together towards a common direction. 
We believe that the effect discussed in this article will be observable in experiments of the kind discussed in Refs.$^{\,}$\cite{nair_science_2008,wang_science_2008,li_natphys_2008,mak_prl_2008}.

Finally let us comment on the broader implications of our results. Effects similar to those described in this article 
are also expected in graphene bilayers and other few-layer systems. 
The lack of Galilean invariance should also affect the cyclotron resonance frequency and oscillator strength when the 2D sheet of graphene is placed in a perpendicular magnetic field.  Undoubtedly much interesting physics, potentially useful for applications in optoelectronics, has still to be learned from the study of graphene and other non-Galilean invariant systems.

\acknowledgments
M.P. acknowledges partial financial support from the CNR-INFM ``Seed Projects" and 
wishes to thank Andrea Tomadin for many invaluable conversations related to the numerical solution of Eq.~(\ref{eq:integral_equation}).
Work in Austin was supported by the Welch Foundation and by NSF Grant No. 0606489.
G.V. acknowledges support from NSF Grant No. 0705460.  
The authors thank Rosario Fazio and Vittorio Pellegrini for a critical reading of the manuscript 
and the FSB for helpful support.


\begin{thebibliography}{77}
\bibitem{electron_doping}
	We discuss electron doping for the sake of definiteness.  
	The graphene properties discussed in this paper are particle-hole symmetric.
\bibitem{cut_off}
	Actually, rigorous gauge invariance is broken by the ultraviolet wavevector cut-off $k_{\rm max}$ and the static pseudospin response 
	is non-zero because the negative energy sea shifts with respect to the circularly symmetric cut-off. 
	This shift, however, does not affect the ``regularized response", which vanishes as expected.
\bibitem{isothermal_adiabatic}
	See Ref.~6 (Chapter 3 and Appendix 8) for other examples  of situations in which the adiabatic and isothermal responses differ.
\end{thebibliography}
\end{document}